# ESTUDO DO ESCOAMENTO E DA DISPERSÃO DE POLUENTES NA SUPERFÍCIE DO LAGO IGAPÓ I

Neyva Maria Lopes Romeiro - Eliandro Rodrigues Cirilo - Paulo Laerte Natti- Universidade Estadual de Londrina - CCE - Departamento de Matemática,
nromeiro@uel.br - ercirilo@uel.br - plnatti@uel.br

**RESUMO**: Ainda são poucas as pesquisas sobre qualidade de água nas cidades, apesar de já existirem muitos levantamentos relativos à poluição de água. Neste sentido o presente trabalho aborda o problema da dispersão de poluentes, através da modelagem matemática e simulação numérica, que tem se mostrado muito eficiente para o estudo de perturbações globais ou mesmo de ecossistemas em particular, como é o caso do Lago Igapó I. O estudo em questão constituiu-se em duas fases. Na primeira modelamos computacionalmente a geometria bidimensional do Lago I, utilizamos a discretização estruturada e em coordenadas generalizadas, sendo a fronteira discreta obtida a partir do método de interpolação polinomial Spline Cúbico Parametrizado. Na seqüência, modelamos matematicamente o escoamento admitindo-o como incompressível, utilizando o sistema de equações de Navier-Stokes e de pressão, submetido a condições iniciais e de contorno, fornecendo os campos de velocidade e pressão em todo o domínio. Inserimos o campo de velocidades na equação de transporte advectiva-difusiva-reativa, nesta segunda fase simulamos numericamente, utilizando o método de elementos finitos, o ciclo do carbono, cujas equações são obtidas através do Modelo QUAL2E. Tendo informações sobre o comportamento da demanda bioquímica de oxigênio e oxigênio dissolvido após um período de tempo, classificamos a qualidade da água do Lago Igapó I conforme os critérios estabelecidos pelo Conselho Nacional do Meio Ambiente (CONAMA), através da Resolução CONAMA 20/86. Pretendemos também, a partir dos resultados, propor medidas locais para melhorar a qualidade da água do lago, via simulação numérica.

**Palavras chaves**: Simulação numérica, Lago Igapó, Demanda bioquímica de oxigênio, Oxigênio dissolvido.

# 1. INTRODUÇÃO

Atualmente diversos setores da sociedade vêm se organizando com o objetivo de frear a degradação do meio ambiente. Muitas alternativas são colocadas em discussão visando à preservação de rios, lagos, mananciais, fundos de vale, matas ciliares, mangues e etc. Cuja existência de forma vigorosa implica na existência da espécie humana. Dentre as várias alternativas, o presente trabalho analisa, por meio de simulação numérica, o impacto que pode causar a descarga de efluentes lançados no Lago Igapó I, localizado em Londrina, Paraná; Brasil. O lago está situado na microbacia do Ribeirão Cambé tendo suas nascentes na cidade de Cambé, aproximadamente 10 km a oeste da cidade de Londrina, no estado do Paraná. Após a nascente, percorre o sentido oeste cruzando toda a porção sul de Londrina, alimentando-se em seu trajeto de vários pequenos córregos. O lago subdivide-se em: Igapó I, Igapó II, Igapó III e Igapó IV. Ele foi projetado em 1957, como uma solução para o problema da drenagem do ribeirão Cambezinho, dificultada por uma barragem natural de pedra. Inicialmente pensou-se em dinamitar a barragem, mas prevaleceu a idéia de formar o lago. O Igapó foi inaugurado em 10 de dezembro de 1959. Ele foi esvaziado e limpo algumas vezes, sendo a última em 1996, onde suas margens foram revitalizadas.

Devido a localização, próximo à região central da cidade de Londrina, e do fato de receber diariamente uma grande quantidade de pessoas que passeiam as suas margens, e outras que praticam esportes náuticos, o Lago Igapó I tornou-se um cartão postal da cidade, por outro lado, o lago também recebe agressão, pois o crescimento da população provocou o aumento do despejo de poluentes não tratados em suas águas. Além disto, existe o despejo clandestino de poluentes nos lagos IV, III e II que polui o Lago I.

Este trabalho encontra-se organizado da seguinte forma: inicialmente é modelada computacionalmente a geometria bidimensional do Lago I. Na sequência, modela-se matematicamente o escoamento admitindo-o como incompressível, fornecendo os campos de velocidade e de pressão em todo o domínio. Obtido o campo de velocidades e inserindo-o na equação de transporte advectiva-difusiva-reativa, pretende-se nesta segunda fase simular numericamente, utilizando o método de elementos finitos [ROMEIRO et al., 2003; ROMEIRO, 2003], o ciclo do carbono, cujas equações são obtidas através do Modelo QUAL2E, obtendo informações sobre o comportamento de duas espécies reativas, sendo a demanda bioquímica de oxigênio e oxigênio dissolvido após um período de tempo. Objetiva-se analisar e classificar a qualidade da água do Lago Igapó I conforme os critérios estabelecidos pelo Conselho Nacional do Meio Ambiente (CONAMA), através da Resolução CONAMA 20/86 [CONAMA, 1986].

## 2. GERAÇÃO DA MALHA COMPUTACIONAL

Neste trabalho a estratégia adotada para a criação da malha computacional do lago foi optar pela discretização estruturada, por que possui vantagens na fácil ordenação dos volumes elementares, o que permite a aplicação de muitos métodos numéricos de solução. E em coordenadas generalizadas, pois este sistema permite que a malha computacional seja coincidente com a geometria do problema, sendo o tratamento computacional mais adequado. Maliska [MALISKA, 1995] afirma que é mais fácil aplicar metodologias numéricas se a discretização for coincidente na fronteira, neste caso os volumes elementares são bem definidos nos contornos.

O domínio físico pode ser visualizado na Figura 1, verificando que a água escoa do lago Igapó II para o Igapó I na passagem pela Avenida Higienópolis, caracterizando a entrada. A margem esquerda é uma região com a presença de vegetação rasteira e caminho para atividade física, no alto existe um canal por onde o lago recebe água. A margem direita é fracionada em propriedades particulares, e como pode ser observado, também existe um canal de recebimento de água pelo lago. A saída constitui-se numa barragem física com passagem controlada de água por adutoras e rampas.

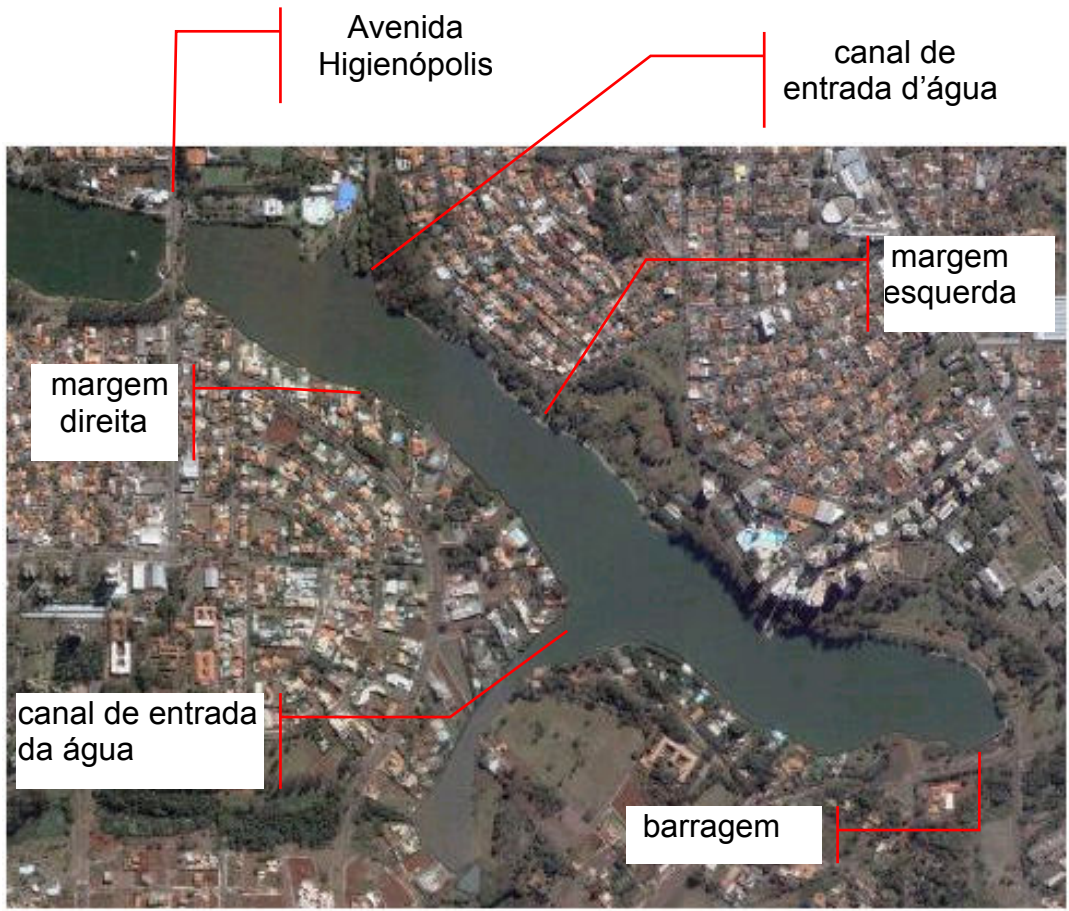

Figura 1: Domínio Físico

A partir de um levantamento aerofotogramétrico, utilizando o sistema de referência não geocêntrico SAD 69 (*South American Datum* 1969), foram obtidas duas amostras de pares ordenados $(x, y)$ das margens esquerda e direita do Igapó I, conforme pode ser observado na Figura 2.

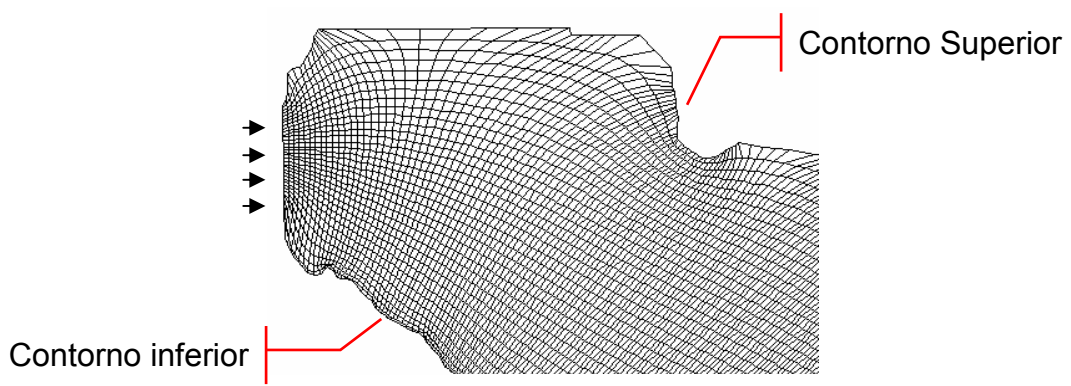

Figura 2: Pontos dos Contornos, Superior e Inferior

A Figura 2 representa um corte da malha do lago na entrada, onde as flechas ilustram a entrada de água. O contorno superior foi obtido com a interpolação dos pontos de sua amostra via o método de interpolação polinomial Spline Cúbico Parametrizado, cujas equações são:

$$x(h) = \sum_{k=0}^{3} a_{k_t}(h-t)^k \qquad (1)$$

e

$$y(h) = \sum_{k=0}^{3} b_{k_t}(h-t)^k \qquad (2)$$

onde a variável $h$ é ajustada na medida que se deseja efetuar, ou não, a concentração de pontos à esquerda ou à direita de um ponto base parametrizado em $t$. Os coeficientes $a_{k_t}$ e $b_{k_t}$ são determinados via resolução de sistemas lineares tridiagonais, o leitor interessado pode consultar [CIRILO, E. R.; BORTOLI, A. L. DE, 2006] para maiores detalhes. De modo análogo obtém-se o contorno inferior, Figura 2.

Utilizamos esse método por que as imposições feitas ao polinômio interpolador minimizam as oscilações [RUGGIERO, M. A. G.; LOPES, V. L. R, 1996]. Além disso, essa técnica é simples do ponto de vista matemático e a sua implementação computacional não é complicada. Dessa forma, o contorno da geometria a ser modelado fica bem representado.

O sistema de EDP's Elíptico, dado por:

$$\alpha x_{\xi\xi} + \gamma x_{\eta\eta} - 2\beta x_{\xi\eta} + \frac{1}{J^2}(P x_\xi + Q x_\eta) = 0 \qquad (3)$$

e

$$\alpha y_{\xi\xi} + \gamma y_{\eta\eta} - 2\beta y_{\xi\eta} + \frac{1}{J^2}(P y_\xi + Q y_\eta) = 0 \qquad (4)$$

permitiram gerar as linhas coordenadas no interior da malha computacional, Figura 2. Em (3) e (4), $x$ e $y$ são as coordenadas cartesianas do domínio físico, $\xi$ e $\eta$ são as coordenadas generalizadas do domínio computacional, $P$ e $Q$ são termos fonte e $J$ é o jacobiano, dado como

$$J = x_\xi y_\eta - x_\eta y_\xi \tag{5}$$

e ainda

$$\alpha = x_\eta^2 + y_\eta^2 \tag{6}$$

$$\beta = x_\eta x_\xi + y_\eta y_\xi \tag{7}$$

$$\gamma = x_\xi^2 + y_\xi^2 \tag{8}$$

com $x_\xi$, $x_\eta$, $y_\xi$ e $y_\eta$ denotando derivadas parciais. O leitor interessado em obter informações de como as grandezas acima são utilizadas para descrever computacionalmente geometrias pode consultar [CIRILO, E. R.; BORTOLI, A. L. DE, 2006] e [DE BORTOLI, A. L., 2000].

Considerando então 839 pontos espaçados ao longo das margens esquerda e direita e 35 pontos espaçados nos contornos de entrada e saída, obtivemos a malha da Figura 3. Note que os canais de entrada d'água foram suprimidos, pois não objetivamos neste trabalho avaliar os gradientes de velocidade da água e as variações de concentração de poluentes nestas regiões.

A metodologia matemática abordada acima permitiu obter a geometria computacional, Figura 3, similar àquela esboçada na Figura 1, que é uma foto de satélite do Lago Igapó I. Esta similaridade é importante porque os modelos hidrodinâmicos e de transporte de poluentes poderão reproduzir soluções numéricas mais fiéis à realidade.

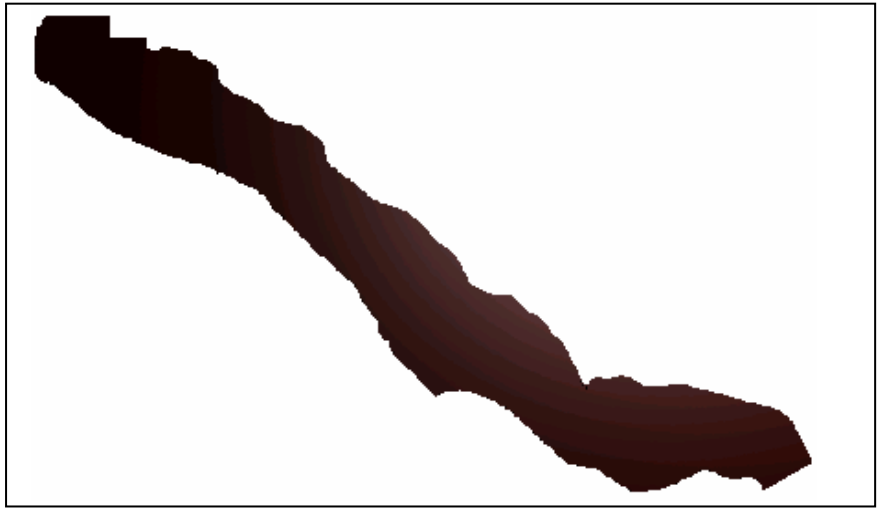

Figura 3: Geometria Computacional

## 3. MODELO HIDRODINÂMICO

A modelagem do problema foi estabelecida segundo algumas hipóteses simplificativas. Consideramos para este trabalho o escoamento em regime permanente, desprezando assim as variações das propriedades com o tempo. Admitimos que as trocas de calor do lago com o meio externo não fossem significativas e que o fluxo ocorresse em temperatura ambiente. Aceitamos o escoamento como incompressível, tal que as reações químicas inerentes à poluição não fossem significativas ao ponto de alterar a massa específica da água, pois do contrário o escoamento deveria ser compressível e o modelo seria outro. Admitimos que os contornos do lago não sofram variações, hipótese do contorno fixo, e que o lago mantivesse sempre a mesma massa de fluido e a mesma área de espelho d'água ao longo do tempo.

Nós consideramos para este trabalho o modelo hidrodinâmico bidimensional, na forma adimensional, constituído pelas leis de conservação:

$$\frac{\partial u}{\partial t} + u\frac{\partial u}{\partial x} + v\frac{\partial u}{\partial y} = -\frac{\partial p}{\partial x} + \frac{1}{\text{Re}}\left(\frac{\partial^2 u}{\partial x^2} + \frac{\partial^2 u}{\partial y^2}\right) \qquad (9)$$

$$\frac{\partial v}{\partial t} + u\frac{\partial v}{\partial x} + v\frac{\partial v}{\partial y} = -\frac{\partial p}{\partial y} + \frac{1}{\text{Re}}\left(\frac{\partial^2 v}{\partial x^2} + \frac{\partial^2 v}{\partial y^2}\right) \qquad (10)$$

$$\nabla^2 p = -\frac{\partial^2 uu}{\partial x^2} - 2\frac{\partial^2 uv}{\partial xy} - \frac{\partial^2 vv}{\partial y^2} - \frac{\partial d}{\partial t} + \frac{1}{\text{Re}}\left(\frac{\partial^2 d}{\partial x^2} + \frac{\partial^2 d}{\partial y^2}\right) \qquad (11)$$

onde $t$ é o tempo, $u$ e $v$ são as componentes do vetor velocidade nas direções $x$ e $y$ respectivamente, Re é o número de Reynolds, $d$ o divergente e $p$ a pressão.

Reescrevemos as equações (9) a (11) para o sistema de coordenadas generalizadas e aproximamos as derivadas parciais por diferenças finitas, obtendo um sistema algébrico de equações. A equação algébrica proveniente de (11), foi resolvida iterativamente através do método das relaxações sucessivas [SMITH, G. D., 1990; DE BORTOLI, A. L., 2000], e desta obtivemos o campo de pressão $p$ para todo o domínio computacional. Inserimos o campo de pressão nas equações algébricas advindas das equações (9) e (10), e resolvemos iterativamente pelo método de Runge-Kutta de terceira ordem [KROLL E ROSSOW, 1989; DE BORTOLI, A. L., 2000], obtivemos as componentes $u$ e $v$ do vetor velocidade em toda a geometria, que por sua vez, foi passado ao modelo de transporte advectivo-difusivo-reativo.

## 4. MODELO DE TRANSPORTE ADVECTIVO-DIFUSIVO-REATIVO

O modelo de reações a ser incluído no modelo de transporte resulta de uma simplificação do modelo linear apresentado por Henze, [HENZE, M.,1987], composto pelo ciclo do carbono, cujas equações são obtidas através do Modelo QUAL2E [CHAPRA, S.C., 1997]. Assim, para analisarmos o impacto da descarga de efluentes lançados no Lago Igapó I, utilizamos um modelo bidimensional de transporte advectivo-difusivo-reativo envolvendo duas espécies reativas, sendo a demanda bioquímica de oxigênio $(X_s)$ e o oxigênio dissolvido $(S_o)$, equações (12) e (13)

$$\frac{\partial X_s}{\partial t} + u\frac{\partial X_s}{\partial x} + v\frac{\partial X_s}{\partial x} - D_x\frac{\partial^2 X_s}{\partial x^2} - D_y\frac{\partial^2 X_s}{\partial y^2} = -(k_1 + k_3)X_s \qquad (12)$$

$$\frac{\partial S_o}{\partial t} + u\frac{\partial S_o}{\partial x} + v\frac{\partial S_o}{\partial x} - D_x\frac{\partial^2 S_o}{\partial x^2} - D_y\frac{\partial^2 S_o}{\partial y^2} = k_2(O_s - S_o) - k_1 X_s \qquad (13)$$

A equação (13) pode ser escrita em termos do déficit de oxigênio dissolvido $(D_{S_o})$, onde $D_{S_o} = O_s - S_o$, sendo $O_s$ a concentração de saturação do oxigênio dissolvido, obtendo-se assim, a seguinte equação diferencial

$$\frac{\partial D_{S_o}}{\partial t} + u\frac{\partial D_{S_o}}{\partial x} + v\frac{\partial D_{S_o}}{\partial x} - D_x\frac{\partial^2 D_{S_o}}{\partial x^2} - D_y\frac{\partial^2 D_{S_o}}{\partial y^2} = k_1 X_S - k_2 D_{S_o} \qquad (14)$$

onde $u$ e $v$ referem-se as componentes do vetor velocidade nas direções $x$ e $y$ respectivamente, gerados pelo modelo hidrodinâmico, $D_x$ e $D_y$ são os coeficiente de difusão nas direções $x$ e $y$ da superfície do lago (estes serão considerados como vetores constante no modelo), $t$ refere-se à variável temporal, $k_1$ é o coeficiente de decomposição da demanda bioquímica de oxigênio, $k_2$ é o coeficiente de reaeração do oxigênio dissolvido, $k_3$ é o coeficiente de sedimentação da demanda bioquímica de oxigênio.

## 5. RESULTADOS NUMÉRICOS

Como já observado, a área a ser avaliada encontra-se definida entre a avenida Higienópolis e a barragem, Figura 3. O ponto de lançamento encontra-se próximo da av. Higienópolis localizado no término do Lago Igapó II e início do Lago Igapó I. O modelo hidrodinâmico, cujo código computacional foi desenvolvido por um dos autores, fundamentado no método das diferenças finitas, nos forneceu o campo de velocidade para todo o domínio, este campo pode ser observado na Figura 4. Ainda, nesta figura, podemos observar as variações das velocidades observando o escala de cores.

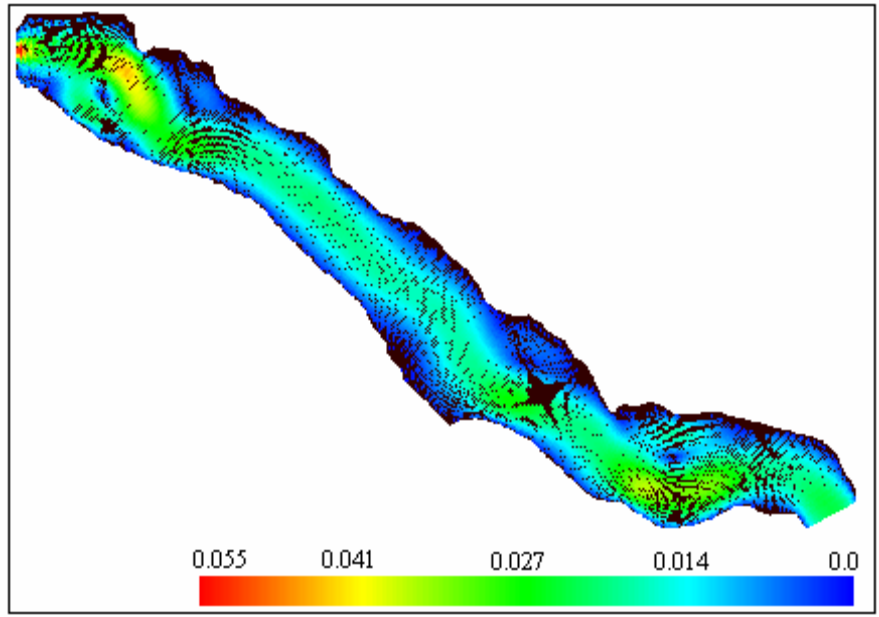

Figura 4: Distribuição do Campo de Velocidade

Para a simulação do modelo descrito em (12) e (14) utilizamos condições iniciais nulas para as duas espécies (demanda bioquímica de oxigênio, déficit de oxigênio dissolvido) , ou seja

$$X_S(x,0) = 0.0 \text{ mgL}^{-1}, \quad D_{S_o}(x,0) = 0.0 \text{ mgL}^{-1} \tag{15}$$

e condições na fronteira

$$X_S(0,t) = 1000.00 \text{ mgL}^{-1}, \quad D_{S_o}(0,t) = 0.0 \text{ mgL}^{-1} \tag{16}$$

O modelo numérico para a simulação das equações (12) e (14) foi resolvido usando o método de elementos finitos estabilizados [BROOKS,1982], onde consideramos uma malha com 23.124 elementos triangulares e 11.886 nós. Utilizamos ainda uma difusão pequena, equivalente a 0.2 $m^2 s^{-1}$ na direção $x$ e zero na direção $y$. A profundidade avaliada $H$ foi de 1.0 (um) metro. Quanto aos valores para os termos lineares de cada uma das espécies, estes foram considerados, $k_1 = 1.157 \times 10^{-7} \, s^{-1}, k_2 = 3.33 \times 10^{-7} \, s^{-1}$ e $k_3 = 9.167 \times 10^{-7} \, s^{-1}$ [ROMEIRO et al., 2002].

Apresentamos nas Figuras 5a-l os resultados das simulações das duas espécies após 458 minutos (aproximadamente sete horas e quarenta minutos) de lançamento contínuo.

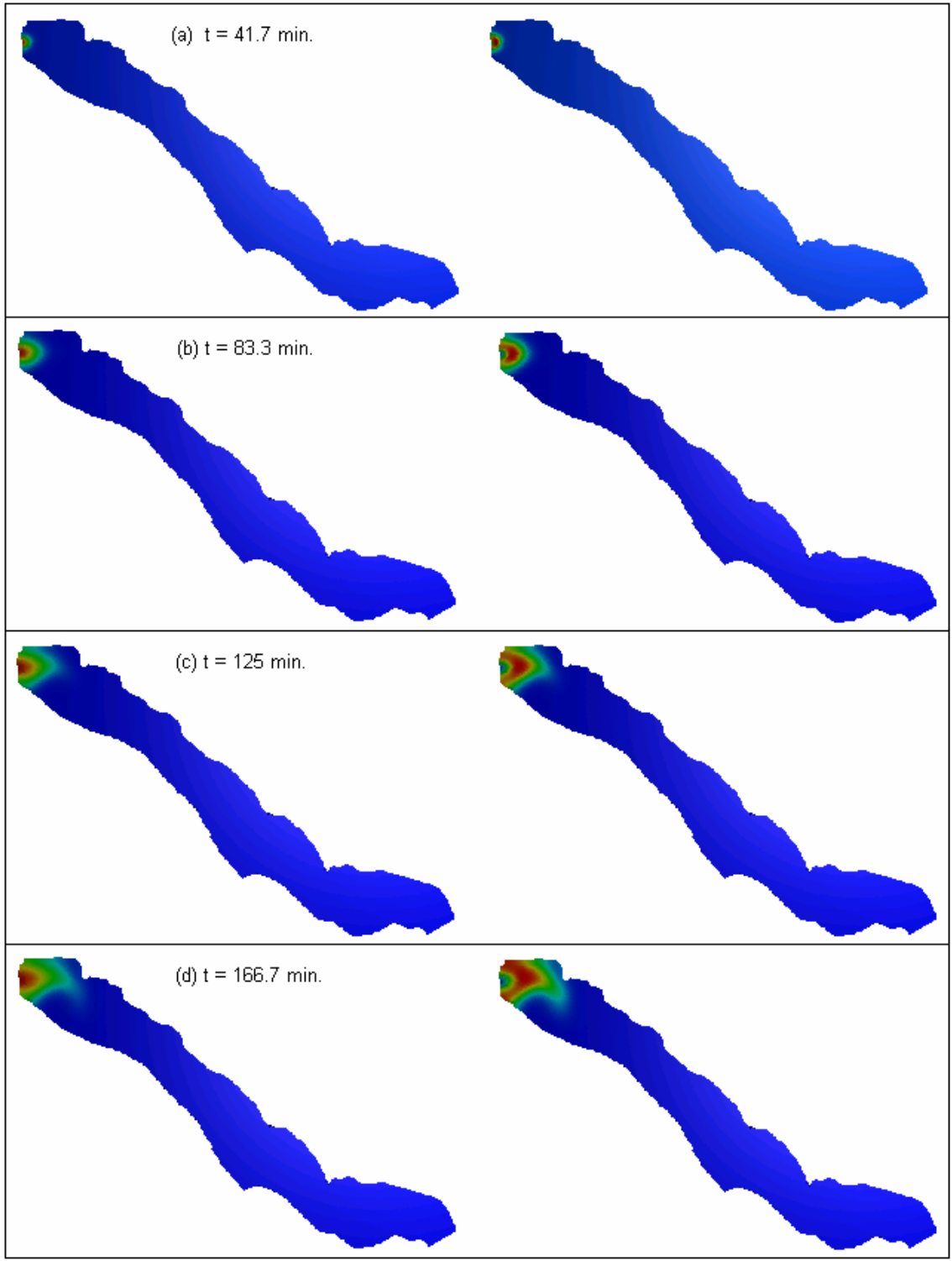

Espécie $X_S$                                Espécie $D_{S_o}$

Figura 5 a)- d): Evolução temporal e espacial das espécies $X_S$ e $D_{S_o}$ considerando um lançamento contínuo

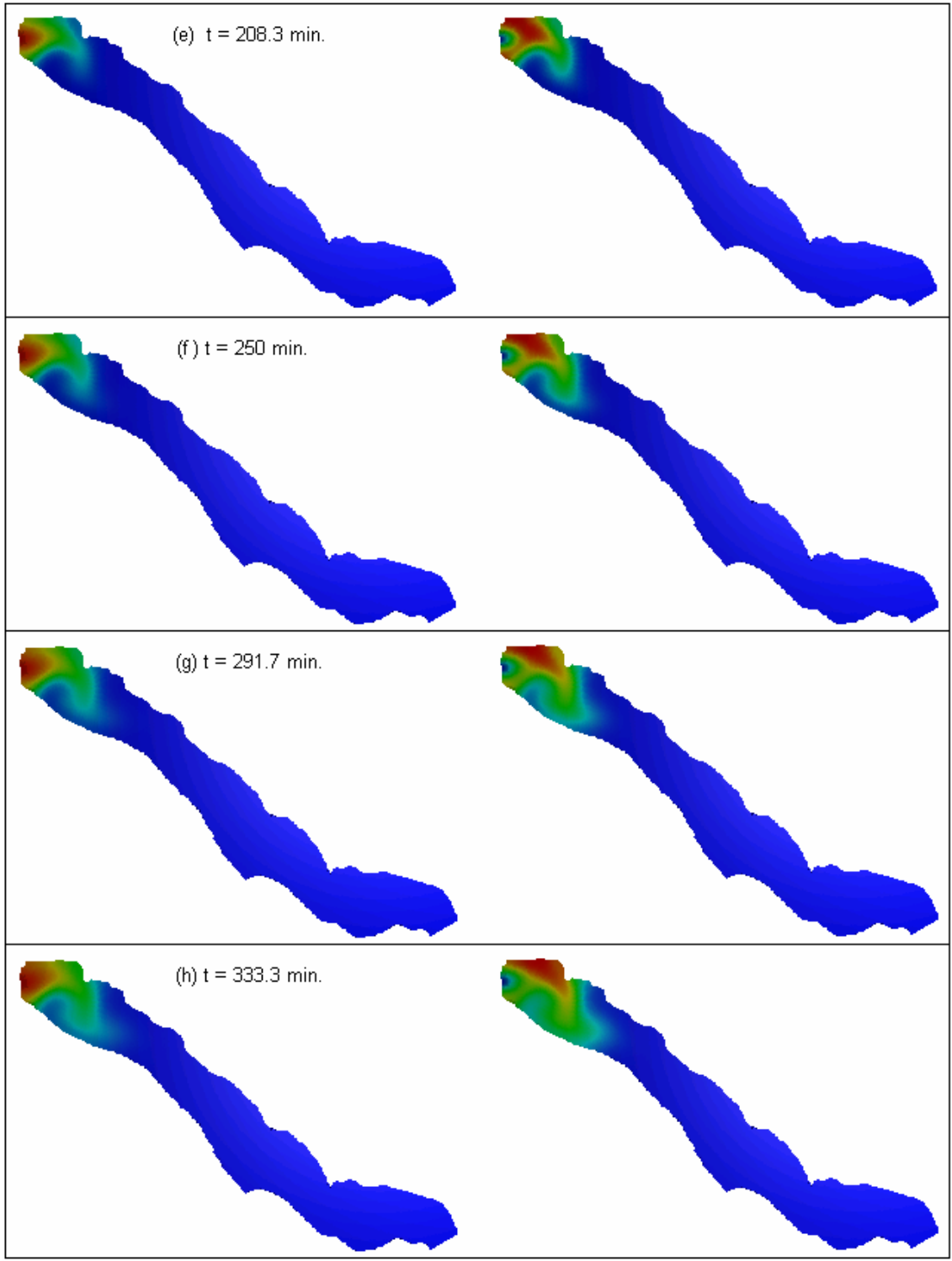

Espécie $X_S$            Espécie $D_{S_o}$

Contin. da Figura 5 e)- h): Evolução temporal e espacial das espécies $X_S$ e $D_{S_o}$ considerando um lançamento contínuo

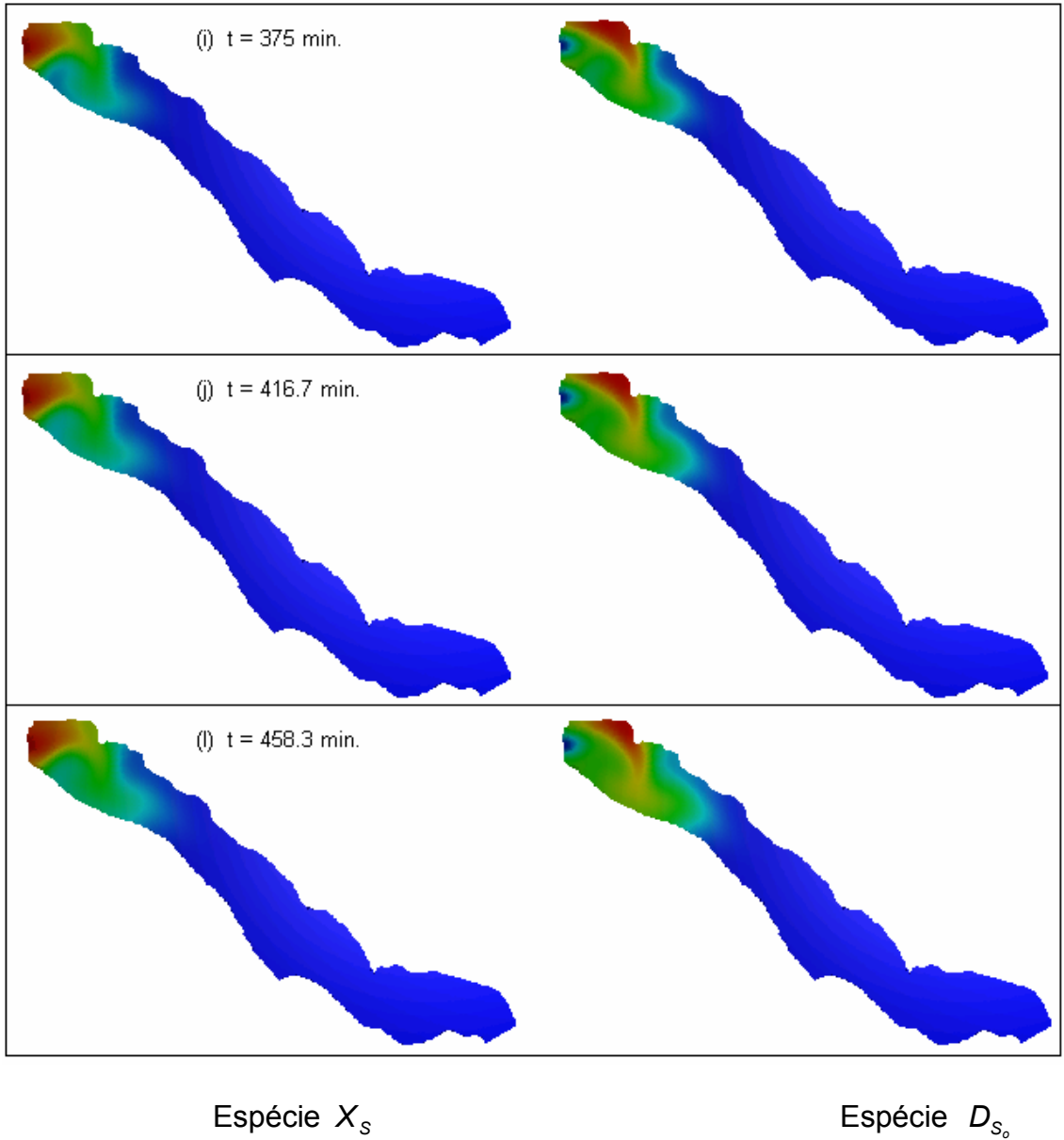

Espécie $X_S$                                     Espécie $D_{S_o}$

Contin. da Figura 5 i)- l): Evolução temporal e espacial das espécies $X_S$ e $D_{S_o}$ considerando um lançamento contínuo superior a sete horas

Observamos nas Figuras 5 a-l variações significativas nas concentrações das espécies simuladas, para aproximadamente um quarto da geometria computacional. Por meio dessas variações, torna-se possível analisar se a descarga lançada está ou não poluindo o Lago Igapó I. Para isso utilizamos os padrões de qualidade de água estabelecidos pela Resolução 20/86 do CONAMA - Conselho Nacional do Meio Ambiente [CONAMA, 1986].

Para evidenciar aquelas variações, apresentamos nas Figuras 6 e 7 o efeito que o lançamento do efluente causa no Lago Igapó I, observando as variações das concentrações por meio das escalas de cores.

O lançando contínuo de uma carga de matéria orgânica correspondente a 1000.00 mg/L (Figura 6) gerou um déficit de oxigênio dissolvido próximo a 2 mg/L (Figura 7b).

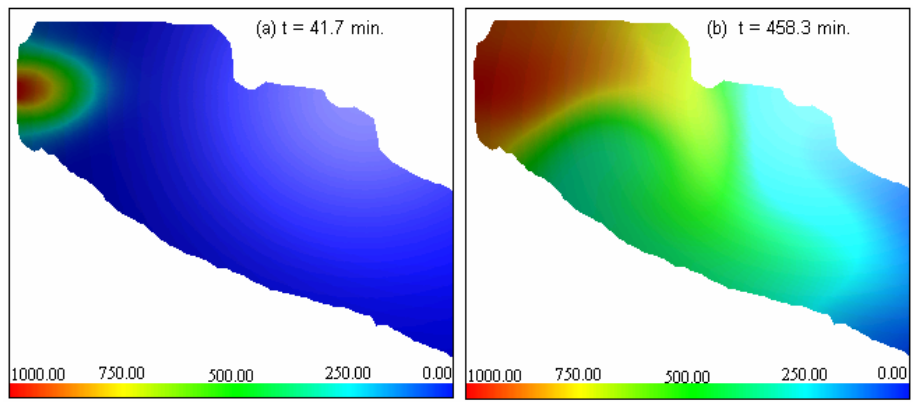

Figura 6 a)- b): Evolução temporal e espacial da demanda bioquímica de oxigênio, $X_S$, considerando um lançamento contínuo: (a) tempo simulado inferior a uma hora, (b) tempo simulado superior a sete horas

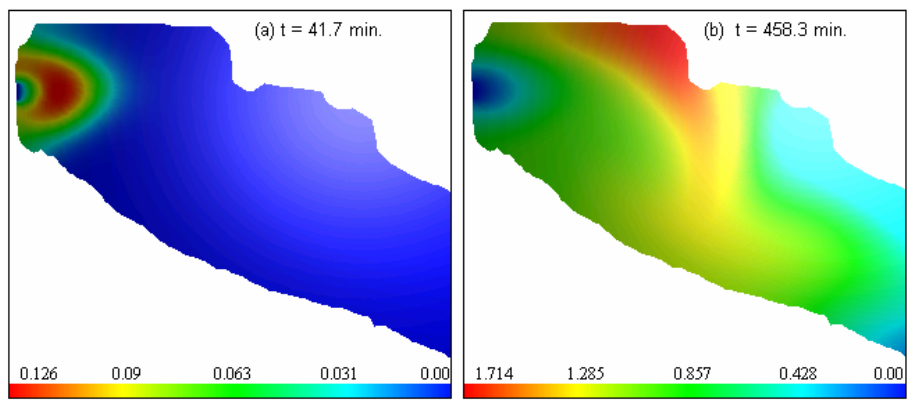

Figura 7 a)- b): Evolução temporal e espacial do déficit de oxigênio dissolvido, $D_{S_o}$, considerando um lançamento contínuo: (a) tempo simulado inferior a uma hora, (b) tempo simulado superior a sete horas

Há de se considerar também que devido a alta concentração da matéria orgânica $(X_S)$ lançada (1000.00 mg/L) obtivemos uma concentração mínima de 250.00 mg/L, não verificando os padrões permitido pela legislação CONAMA inferior a 5 mg/L – referente a condições adequadas para abastecimento doméstico (após um tratamento convencional) e recreação de contato primário, ou inferior a 3 mg/L referente a características desejadas para pesca e preservação das comunidades aquáticas – como pode ser observado na Figura 6.

Quanto ao efeito causado pelo lançamento da matéria orgânica que diminui a concentração do oxigênio dissolvido no lago, este pode ser observado pelos resultados simulados via o déficit de oxigênio dissolvido $(D_{S_o})$. Para as simulações, consideramos a concentração de saturação do oxigênio dissolvido em $O_s = 8.0$ mg/L. Disso, via $S_o = O_s - D_{S_o}$, obtivemos que a maior concentração (no tempo simulado) de $D_{S_o}$ foi de 1.714 mg/L na margem esquerda (Figura 7b), aqui a concentração de $S_o$ foi próxima de 6 mg/L. Este fato ocorreu devido à presença do vórtice, Figura 8.

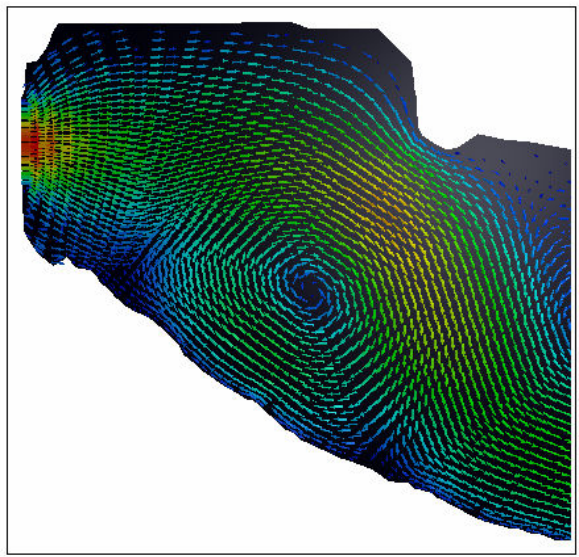

Figura 8: Vórtice na entrada do lago

**CONCLUSÃO**

No presente trabalho analisamos, por meio da simulação numérica, o impacto causado por uma descarga de efluentes lançados no Lago Igapó I, localizado em Londrina, Paraná, Brasil. Para esta simulação usamos um modelo de qualidade de água composto pelos modelos hidrodinâmico e de transporte advectivo-difusivo-reativo.

Concluímos por meio das simulações que uma alta carga lançada no lago pode resultar em sérios problemas tanto para a população humana quanto para a aquática, como pode ser observado nas Figuras 6 e 7. Com o fato de que a margem esquerda apresentou uma concentração maior que a da direita, devido ao fato da presença do vórtice naquela região, Figura 8.

Com o desenvolvimento deste trabalho, a técnica como já abordada, inova por causa da sua simplicidade matemática, computacional e eficiência. O contorno pode ser modelado por Spline Parametrizado a partir de uma quantidade pequena de pontos conhecidos. Os modelos hidrodinâmico e de transporte foram úteis para descrever os gradientes de velocidade, concentrações da demanda bioquímica de oxigênio e do déficit de oxigênio dissolvido.

O próximo passo será tratar da poluição em toda a geometria, simulando vários valores para a descarga de efluentes em localizações cujo fluxo de água, calculado pelo modelo hidrodinâmico, esteja livre. Porque o nosso objetivo é diminuir o impacto da alta concentração de poluição, cuja intenção deste trabalho encontra-se direcionada.

**REFERÊNCIAS**

BROOKS, A.N., HUGHES,T.J. R., *Streamline upwind/Petrov-Galerkin formulation for convection dominated flows with particular emphasis on the incompressible Navier-Stokes equations*, Computer Methods in Applied Mechanics and Engineering, v. 32, pp. 199-259, 1982.

CHAPRA, S.C., 1997, *Surface Water Quality Modeling*. McGraw-Hill International.

CIRILO, E. R.; BORTOLI, A. L. DE., *Geração do Grid da Traquéia e dos Tubos Bronquiais por Splines Cúbicos*. Semina. Ciências Exatas e Tecnológicas, v. 27, pp. 147-155, 2006.

CONAMA, *Resolução nº 20, do Conselho Nacional do Meio Ambiente, de 18/06/86, Legislação Federal de Controle da Poluição Ambiental*. Série Documentos ISSN 0103-264X, CETESB Companhia de Tecnologia de Saneamento Ambiental, Secretaria do Meio Ambiente, São Paulo, SP, Brasil, 1995.

DE BORTOLI, A. L., *Introdução à Dinâmica de Fluidos Computacional*. Ed. da UFRGS, Porto Alegre-RS. 2000.

HENZE, M., GRADY C. P. L., GUJER et al. - *Activated Sludge Model*, no 1., IAWQ, London, 1987.

MALISKA, C. R., *Transferência de Calor e Mecânica dos Fluidos Computacional, Fundamentos e Coordenadas Generalizadas*, LTC- Livros Técnicos e Científicos Editora S.A, Rio de Janeiro – RJ. 1995.

ROMEIRO, N.M.L, CASTRO, R.G.S. E LANDAU, L. (2003). Simulação numérica do transporte de poluentes envolvendo quatro espécies reativas nos Rios Negro e Solimões. I Simpósio de Recurso Hídricos da Amazônia, Manaus-AM, Brasil. Caderno de resumos pp. 50 e em CD-ROM.

ROMEIRO, N.M.L, RIBEIRO, F.L.B. E LANDAU, L. (2002). *A finite element formulation for shallow water flow and reactive pollutant transport*. CILAMCE, 2001, Rio de Janeiro-RJ. CD-ROM.

ROMEIRO, N.M.L. (2003), Simulação Numérica de Modelos de Qualidade de água usando o Método de Elementos Finitos Estabilizados, *Tese de Doutorado*, COPPE/UFRJ/RJ, Programa de Engenharia Civil.

RUGGIERO, M. A. G.; LOPES, V. L. R., *Cálculo numérico: aspectos teóricos e computacionais.* São Paulo: Makron, 1996.

SHAKIB, F. (1989), *Finite Element Analysis of the Compressible Euler and Navier-*

SMITH, G. D., *Numerical solution of partial differential equations: finite difference methods.* New York: Oxford University Press, 1990.